\documentclass[twocolumn,showpacs,prl,amsmath,amssymb]{revtex4}

\usepackage{graphicx}% Include figure files
\usepackage{dcolumn}% Align table columns on decimal point
\usepackage{bm}% bold math
\usepackage{amsmath,epsfig}

\newcommand{\ket}[1]{\left\vert#1\right\rangle}
\newcommand{\bra}[1]{\left\langle#1\right\vert}
\newcommand{\braket}[2]{\left\langle#1\vert#2\right\rangle}

\begin{document}

%\preprint{APS/123-QED}

\title{On the observability of Bell's inequality violation in the optical Stern-Gerlach model}

\author{M. Tumminello \dag, A. Vaglica \ddag, G. Vetri \ddag}
\affiliation{%
\dag Istituto Nazionale di Fisica della Materia and
Dipartimento di Fisica e Tecnologie Relative, Universit\`{a} degli Studi di Palermo,
Viale delle Scienze, Palermo, I-90128, Italy\\
\ddag Istituto Nazionale di Fisica della Materia and Dipartimento di Scienze Fisiche ed Astronomiche,Universit\`{a}
degli Studi di Palermo,
via Archirafi 36, 90123 Palermo, Italy
}%

\date{\today}% It is always \today, today,
             %  but any date may be explicitly specified

\begin{abstract}
Using the optical Stern-Gerlach model, we have recently shown that
the non-local correlations between the internal variables of two
atoms that successively interact with the field of an ideal cavity
in proximity of a nodal region are affected by the atomic
translational dynamics. As a consequence, there can be some
difficulties in observing violation of the Bell's inequality for
the atomic internal variables. These difficulties persist even if
the atoms travel an antinodal \textit{region}, except when the
spatial wave packets are exactly
centered in an antinodal \textit{point}.\\
\end{abstract}

\maketitle

Peculiar concepts of quantum mechanics (QM), such as the Bohr's
principle of complementarity \cite{Scu, Zei, Durr, Bert}, have
their origin in the vectorial nature of the state space, which
involves a superposition principle. Complementarity (or duality)
\cite{Engl} establishes a sort of ``orthogonality" between the
which-way information and the possibility of observing
interference pattern. In other words, these two behaviors are
mutually exclusive. The visibility $V$ of the interference pattern
and the distinguishability $D$ of the quantum paths can in some
extent coexist, and as shown by Englert in its quantitative
analysis of complementarity \cite{Engl,See}, they satisfy the
inequality $D^{2}+V^{2}\leq1$. According to this analysis and in
the ambit of the optical Stern-Gerlach (SG) model, we have
recently shown \cite{Tum1} that the visibility of the Rabi
oscillations and the distinguishability of the two atomic
translational paths satisfy the equality relation $D^{2}+V^{2}=1$
when pure initial states are considered.

When applied to a composite system, the superposition principle
leads to quantum correlations (entanglement), which may hide the
individuality of the subsystems. Differently from the classical
case, and in idealized configurations, two quantum systems that
have interacted for a time, generally do not recover their
individuality, even if the subsystems become spatially separated.
This inseparability, which has been at the origin of the famous
debate between Einstein \cite{Ein} and Bohr \cite{Boh} on the
completeness of QM, implies a non local character of the
correlations (EPR correlation \cite{Wer}) between the two
subsystems. This non-locality can be individuated by the violation
of some Bell's inequality \cite{Bel}.
It is to note that, differently from the pure case, a mixed state
may be EPR correlated and, at the same time, it may satisfy the
Bell's inequality \cite{Wer}.

Recently, it has been payed attention to teleportation, non-local
correlations, separability and related issues for massive
particles \cite{Har, Bar, Rie, Phoe, Kni, Fre, Zub, Ray, Tum2}. As
suggested by Phoenix and Barnett \cite{Phoe} (see also \cite{Har}
and \cite{Kni}), a simple model which can realize an EPR state for
massive particles consists of two atoms which interact
successively with the field of an optical cavity, in the ambit of
the standard Jaynes-Cummings (JC) model. The entanglement
developed during the interaction between the first atom and the
field, may induce quantum correlations between the two atoms as
the second one interacts with the field of the same cavity. An
experimental effort to observe a Bell's inequality violation for
this system has been done by Haroche and co-workers \cite{Har},
which ascribe to several experimental imperfections the reduction
of purity of the entanglement that prevents the Bell's inequality
violation.

In a recent paper \cite{Tum2} we have suggested that a careful
analysis of the interatomic correlations may require the
quantization of the translational dynamics of the two atoms along
the cavity axis. In that paper we have considered two-level atoms
entering the cavity in a nodal region, where the field gradient is
different from zero. The entanglement between the internal and
external atomic variables affects the non local features of the
interatomic correlations making more difficult the observation of
Bell's inequality violation with respect to the JC model. However,
in most cases (as in ref.: \cite{Har}) the experiments are performed in such a way that the
atoms interact with the cavity field in an antinodal region.
Consequently, it seems suitable to extend our previous analysis to
this case. Our present analysis  confirms qualitatively the
results of the previous one, except when the spatial atomic wave
packets are exactly centered in an antinodal \textit{point}.

In our model two two-level atoms interact successively with a
single mode of the e.m. field of an ideal cavity. The first atom,
say $A_{1}$, enters the cavity at time $t=0$ and interacts with
the field for a time $t_{1}$. It moves prevalently along the
\emph{z}-direction, orthogonal to the  \emph{x}-cavity axis and we
assume that the velocity along this direction is large enough to
treat classically this component of the motion. The second atom,
say $A_{2}$, enters the cavity at time $t_{2}>t_{1}$, interacts
with the e.m. field as modified by the first atom and leaves the
cavity at time $t_{3}$. Finally, both the atoms evolve freely for
$t>t_{3}$. The atoms enter the cavity in proximity of an antinodal
region of the resonant \emph{k}-mode, and the width of their wave
packets is sufficiently small with respect to the wavelength of
this mode. The Hamiltonian of the system at all times can conseguently be written as

\begin{widetext}
\begin{equation}\label{ham}
  \hat{H}=\frac{\hat{p}_{1}^{2}}{2\, m}+\hbar \omega (\hat{a}^{\dag}\hat{a}+\hat{S}_{z,1}+
  \frac{1}{2})+(\frac{\hat{p}_{2}^{2}}{2\,m}+\hbar \omega \hat{S}_{z,2})\theta_{t}(t_{2})+
  \hbar\varepsilon (\frac{k^{2}\,\hat{x}_{1}^{2}}{2}-1)\mu_{t}(0,t_{1})\hat{u}_{1}
  +\hbar\varepsilon (\frac{k^{2}\,\hat{x}_{2}^{2}}{2}-1)\mu_{t}(t_{2},t_{3})\hat{u}_{2},
\end{equation}
\end{widetext}
where $\hat{x}_{i}$ is the position of atom $A_{i}$ with respect
to the antinodal point and $\hat{p}_{i}$ its conjugate momentum.
The atom-field interaction is described by
$\hat{u}_{i}=\hat{a}^{\dag}\hat{S}_{-,i}+\hat{a}\hat{S}_{+,i}$
where $\hat{a}$ and $\hat{a}^{\dag}$ are the usual annihilation
and creation field-operators, while $\hat{S}_{\pm,i}$ are the
$1/2$ spin operators. The atoms have same mass \emph{m} and same
atom-field coupling constant $\varepsilon$. The linear combination
of step-functions $\mu_{t}(x,y)=\theta_{t}(x)-\theta_{t}(y)$, with
different points (x and y) of discontinuity, distinguishes the
different time ranges concerning the successive interactions.

As in Ref. \cite{Phoe} where the standard JC model is adopted and
as in our previous paper \cite{Tum2}, we consider the simple case
of only one atom-field system excitation. In particular, we start
considering both the atoms initially in the ground state and just
one photon in the cavity, so at time $t=0$ the state is
$\ket{\psi(0)}=\ket{g_{1}}\ket{1}\ket{\varphi_{1}(0)}$,
where $\ket{\varphi_{1}(0)}$ is a translational state of the atom
$A_{1}$.

Using the evolution operator related to eq.(\ref{ham}), the state
of the system for $t\leq{t_{2}}$ is (except an irrelevant global
phase factor)
\begin{eqnarray}\label{psit}
\ket{\psi(t)}=exp[-\frac{i}{\hbar}\frac{\hat{p}_{1}^{2}}{2 m}
(t-t_{1})]\cdot\quad\quad\quad\quad\quad\quad\quad\nonumber\\
\quad\cdot[\ket{S_{1}^{-}(t_{1})}\ket{e_{1}}\ket{0}\
+\ket{S_{1}^{+}(t_{1})}\ket{g_{1}}\ket{1}],
\end{eqnarray}
where  $\ket{e_{i}}$ indicates the excited state of $A_{i}$ and
\begin{eqnarray}
\ket{S_{1}^{\pm}(t_{1})}=\frac{1}{2}[e^{{i}\varepsilon
t_{1}}\ket{\phi_{1}^{+}(t_{1})}\pm e^{{-i}\varepsilon
t_{1}}\ket{\phi_{1}^{-}(t_{1})}] \label{Spm1}\\
\ket{\phi_{1}^{\pm}(t_{1})}=
\exp\{-\frac{i}{\hbar}[\frac{\hat{p}_{1}^{2}}{2 m}\pm\hbar
\varepsilon
k^{2}\frac{\hat{x}_{1}^{2}}{2}]t_{1}\}\ket{\varphi_{1}(0)}.
\label{fipm1t1}
\end{eqnarray}
At time $t=t_{2}$ the second atom, in its ground internal state,
enters the cavity and starts to interact with the field modified
by the interaction with the first atom. Let
$\ket{\varphi_{2}(t_{2})}$ be the translational state of atom
$A_{2}$ at the beginning of its interaction with the cavity field.
The state of the entire system at this time is
$\ket{\Psi(t_{2})}=\ket{\psi(t_{2})}\ket{g_{2}}\ket{\varphi_{2}(t_{2})}$.
Applying the same procedure as above, we derive the state at time
$t>t_{3}$, when both the two atoms have left the cavity and evolve
freely
\begin{eqnarray}\label{Psit}
\ket{\Psi(t)}=\ket{S_{1}^{+}(t)}\ket{S_{2}^{+}(t)}\ket{g_{1}}\ket{g_{2}}\ket{1}+\quad\quad\quad
\quad\quad\quad\nonumber\\
+\{\ket{S_{1}^{-}(t)}\ket{\varphi_{2}(t)}\ket{e_{1}}\ket{g_{2}}+\quad\quad\quad\quad
\quad\quad\quad\nonumber\\
+\ket{S_{1}^{+}(t)}\ket{S_{2}^{-}(t)}\ket{g_{1}}\ket{e_{2}}\}\ket{0}\quad\quad
\end{eqnarray}
where
\begin{eqnarray}
\ket{S_{i}^{\pm}(t)}=exp[-\frac{i}{\hbar}\frac{\hat{p}_{i}^{2}}{2
m}(t-t_{3})]\ket{S_{i}^{\pm}(t_{3})}\quad\label{St}\\
\ket{S_{i}^{\pm}(t_{3})}=\frac{1}{2}[e^{{i}\varepsilon
T_{i}}\ket{\phi_{i}^{+}(t_{3})}\pm e^{{-i}\varepsilon T_{i}}
\ket{\phi_{i}^{-}(t_{3})}]\quad\label{St3}\\
\ket{\phi_{1}^{\pm}(t_{3})}=exp[-\frac{i}{\hbar}\frac{\hat{p}_{1}^{2}}{2
m}(t_{3}-t_{1})]\ket{\phi_{1}^{\pm}(t_{1})}\quad
\label{fipm1t3}\\
\ket{\phi_{2}^{\pm}(t_{3})}=\exp\{-\frac{i}{\hbar}[\frac{\hat{p}_{2}^{2}}{2
m}\pm\hbar \varepsilon
k^{2}\frac{\hat{x}_{2}^{2}}{2}]T_{2}\}\ket{\varphi_{2}(t_{2})}
\label{fipm2t3}\\
\ket{\varphi_{2}(t_{3})}=exp[-\frac{i}{\hbar}\frac{\hat{p}_{2}^{2}}{2
m}T_{2}]\ket{\varphi_{2}(t_{2})},\label{fi2t3}\quad
\end{eqnarray}
and we have introduced the interaction time $T_{1}=t_{1}$ and
$T_{2}=t_{3}-t_{2}$ for atoms $A_{1}$ and  $A_{2}$, respectively.
Tracing on the field and atomic translational variables, the
following reduced density operator is obtained
\begin{widetext}
\begin{eqnarray}\label{rhorid}
\rho=Tr_{f,s_{1},s_{2}}(\ket{\Psi(t_{3})}\bra{\Psi(t_{3})})=
\frac{1}{4}(1+c_{R}^{(1)})(1+c_{R}^{(2)})\ket{g_{1}}\ket{g_{2}}\bra{g_{1}}
\bra{g_{2}}+\frac{1}{2}(1-c_{R}^{(1)})\ket{e_{1}}\ket{g_{2}}\bra{e_{1}}
\bra{g_{2}}+ \nonumber \\
+\frac{1}{4}(1+c_{R}^{(1)})(1-c_{R}^{(2)})\ket{g_{1}}\ket{e_{2}}\bra{g_{1}}
\bra{e_{2}}+ \frac{i}{4}
c_{I}^{(1)}[(c_{-}-c_{+})\ket{e_{1}}\ket{g_{2}}\bra{g_{1}}
\bra{e_{2}}-h.c.],
\end{eqnarray}
\end{widetext}
where we have put
\begin{eqnarray}\label{param}
e^{{-2i}\varepsilon
T_{1}}\braket{\phi_{1}^{+}(t_{1})}{\phi_{1}^{-}(t_{1})}=c_{R}^{(1)}+ic_{I}^{(1)}\nonumber\\
e^{{-2i}\varepsilon
T_{2}}\braket{\phi_{2}^{+}(t_{3})}{\phi_{2}^{-}(t_{3})}=c_{R}^{(2)}+ic_{I}^{(2)}\nonumber\\
e^{\mp{i}\varepsilon
T_{1}}\braket{\phi_{2}^{\pm}(t_{3})}{\varphi_{2}(t_{3})}=c_{\pm}.\quad\quad\quad
\end{eqnarray}
As it is easy to see, eq. (\ref{rhorid}) is formally very similar
to the corresponding equation of Ref.\cite{Phoe}. The difference
between eq. (\ref{rhorid}) and the corresponding in
Ref.\cite{Phoe} is the fact that the coefficients (\ref{param})
are now affected by the translational dynamics. As in
Ref.\cite{Tum2}, the scalar products which appear in this equation
are generally subjected to a non dissipative decay and this
behavior may affect the non local character of the correlations
between the internal atomic variables.

An evaluation of the quantities (\ref{param}) is not a trivial
operation because the evolution operator in eq.s (\ref{fipm1t1}) and
(\ref{fipm2t3}) describes a harmonic-like evolution (sign $-$) or a squeezing-like evolution
in the other case (sign $+$) \cite{Vag}. In fact one can write
\begin{equation}\label{sq}
e^{-\frac{i}{\hbar}\left(\frac{\hat{p}_{i}^{2}}{2\,m}- \frac{\hbar
\varepsilon k^{2}}{2}\,\hat{x}_{i}^{2}\right)\,T_{i}}=
e^{i\,\frac{\omega_{0}}{2}\,T_{i}\left(\hat{b}_{i}^{\dag\,2}+\hat{b}_{i}^{2}\right)},\
\end{equation}
where $T_{i}$ is the interaction time of atom $A_{i}$,
\begin{equation}\label{bosons}
\hat{b}_{j}=\frac{1}{\sqrt{2}}\left(\sqrt{\frac{m\,\omega_{0}}{\hbar}}\,\hat{x}_{j}+
i\,\frac{1}{\sqrt{m\,\omega_{0}\hbar}}\,\hat{p}_{j}\right)
\end{equation}
are boson operators and $\omega_{0}^2 =\frac{\hbar k^{2}}{m}
\varepsilon$. To calculate the scalar products (\ref{param}), it
is convenient to put the squeezing operators (\ref{sq}) in
factored forms \cite{cav}, for instance
\begin{eqnarray}\label{fact}
e^{i\,\frac{\alpha}{2}\left(\hat{b}_{i}^{\dag\,2}+\hat{b}_{i}^{2}\right)}=
\exp\{-\ln[\cosh(\alpha)](\hat{b}_{i}^\dag\
\hat{b}_{i}+\frac{1}{2})\}\cdot\nonumber\\
\cdot\exp\{\frac{i}{2}\tanh(\alpha)e^{2\ln[\cosh(\alpha)]}
\hat{b}_{i}^{\dag\,2}\}
\cdot\exp\{\frac{i}{2}\tanh(\alpha)\hat{b}_{i}^{2}\},
\end{eqnarray}
and similar expressions. Moreover, for the sake of mathematical
simplicity, we assume that the initial translational states for
both the atoms are given by coherent states of the boson-like
operators $b_{j}$, with the same width. In other words, we suppose
that at the beginning of the interaction with the cavity field the
translational states of both the atoms are coherent states with
respect to the bosons operators $\hat{b}_{j}$:
$\ket{\varphi_{j}(initial)}$=$\ket{\alpha_{j}}$, with
$\hat{b}_{j}\ket{\alpha_{j}}=\alpha_{j}\ket{\alpha_{j}}$, and
\begin{eqnarray}\label{psinit}
\ket{\alpha_{j}}=\exp\left[\frac{i}{\hbar}\left(p_{0}^{(j)}
\hat{x}_{j}-x_{0}^{(j)}\hat{p}_{j}\right)\right]\ket{0_{j}}\\
\alpha_{j}=x_{0}^{(j)}\sqrt{\frac{m\omega_{0}}{2\hbar}}+ip_{0}^{(j)}\frac{1}{\sqrt{2m\hbar\omega_{0}}}
\equiv a_{j}+ib_{j}
\end{eqnarray}
where
\begin{equation}\label{psivac}
\braket{x_{j}}{0_{j}}=\left(\frac{1}{\Delta
x_{0}\sqrt{2\,\pi}}\right)^{\frac{1}{2}}\,\exp[-\frac{x_{j}^{2}}{4\,\Delta
x_{0}^{2}}]\\
\end{equation}
is the wave function of the ground state of the $\hat{b}_{j}$
corresponding harmonic ``oscillator" and $\Delta
x_{0}^2=\frac{\hbar}{2\,m\,\omega_{0}}$ is the same for both the
atoms. This choice is not too restrictive because the only
restriction introduced with respect to a minimum uncertainty
gaussian packet with arbitrary initial momentum $p_{0}$ and
position $x_{0}$ is its wideness. Furthermore, it is to notice
that a general packet can always be expressed as a superposition
of coherent states. Using eq.s (\ref{fact}) and (\ref{psinit}),
the scalar products which appear in eq. (\ref{param}) assume the
following form,
\begin{widetext}
\begin{eqnarray}
\braket{\phi_{1}^{+}(T_{1})}{\phi_{1}^{-}(T_{1})}=\bra{\alpha_{1}}e^{\{i\,\omega_{0}\,T_{1}-\ln[\cosh(\omega_{0}
T_{1})]\}\left(\hat{b}_{1}^{\dag}\,\hat{b}_{1}+\frac{1}{2}\right)}\,e^{\frac{i}{4}\sinh(2\,\omega_{0}\,T_{1})\,\hat{b}_{1}^{\dag\,2}}\,
e^{\frac{i}{2}\tanh(\omega_{0}\,T_{1})\,\hat{b}_{1}^{2}}\ket{\alpha_{1}}\label{phiupphidownboson}\\
\braket{\phi_{2}^{+}(T_{2})}{\varphi_{2}(T_{2})}=\bra{\alpha_{2}}e^{i\,\omega_{0}\,T_{2}\,
\left(\hat{b}_{2}^{\dag}\,\hat{b}_{2}+\frac{1}{2}\right)}\,e^{-i\,\frac{\hat{p}_{2}^{2}}{2\,m\,\hbar}\,T_{2}}
\ket{\alpha_{2}}\label{phiupphi0boson}\\
\braket{\phi_{2}^{-}(T_{2})}{\varphi_{2}(T_{2})}=\bra{\alpha_{2}}
e^{-\frac{i}{2}\tanh(\omega_{0}\,T_{2})\,\hat{b}_{2}^{\dag\,2}}\,
e^{-\frac{i}{4}\sinh(2\,\omega_{0}\,T_{2})\,\hat{b}_{2}^{2}}\,
e^{\{-\ln[\cosh(\omega_{0}T_{2})]\}\left(\hat{b}_{2}^{\dag}\,\hat{b}_{2}+\frac{1}{2}\right)}\,
e^{-i\,\frac{\hat{p}_{2}^{2}}{2\,m\,\hbar}\,T_{2}}\,
\ket{\alpha_{2}}\label{phidownphi0boson}
\end{eqnarray}
\end{widetext}
A straightforward calculation leads now to the evaluation of these
scalar products, where the expansion of the state
$\exp[-i\,\frac{\hat{p}_{2}^{2}}{2\,m\,\hbar}\,t]\,
\ket{\alpha_{2}}$ in terms of coherent states corresponding to
the second atom boson-like operators is required.\\
For $x_{0} \neq 0$ and/or $p_{0} \neq 0$ these terms are
characterized by a non dissipative damping. For example, the
scalar product (\ref{phiupphidownboson}) for $t\leq{T_{1}}$
behaves as
\begin{widetext}
\begin{eqnarray}
\braket{\phi_{1}^{+}(t)}{\phi_{1}^{-}(t)}=e^{i\frac{\omega_{0}}{2}t}e^{-i|\alpha_{1}|^2\frac{\sin(\omega_{0}t)}{\cosh(\omega_{0}t)}}
\exp\{\frac{i}{2}\tanh(\omega_{0}t)[(a_{1}^2-b_{1}^2)(1+\cos(2\omega_{0}t))+2a_{1}b_{1}\sin(2\omega_{0}t)]\}\cdot\nonumber\\
\cdot\frac{1}{\sqrt{\cosh(\omega_{0}t)}}e^{-|\alpha_{1}|^2(1-\frac{\cos(\omega_{0}t)}{\cosh(\omega_{0}t)})}
\exp\{-\tanh(\omega_{0}t)[a_{1}b_{1}(1-\cos(2\omega_{0}t))+\frac{1}{2}(a_{1}^2-b_{1}^2)\sin(2\omega_{0}t)]\}\label{phiupdown1}\\
\propto{[1-\frac{(\omega_{0}t)^2}{2}]}\cdot\exp{\{-2a_{1}^2(\omega_{0}t)^2\}}\label{phiupdown1red}\quad\quad\quad\quad\quad\quad\quad
(\omega_{0}t<1)\quad\quad
\quad\quad\quad\quad\quad\quad\quad\quad\quad\quad\quad
\end{eqnarray}
\end{widetext}
The damping factor shown by this last approximated expression,
which is at the origin of the non dissipative damping of the Rabi
oscillations \cite{Vag,Cus,Tum1}, is due to the increasing
distance in the phase space \cite{Chian} of the two deflected
components of the translational wave packet \cite{Aha}. Similar
behaviors hold for the other coefficients of eq.(\ref{param}). The
condition $(\omega_{0}t<1)$ is not much restrictive for the
parameters used in this paper, and at the same time, it is in
agreement with the quadratic approximation of the cavity mode
function. After a few periods of Rabi oscillations, the damping
factors involved in the scalar products (\ref{param}) when $x_0
\neq 0$, determine a decoherence of the system described by the
density matrix (\ref{rhorid}), i.e. the last term in eq.
(\ref{rhorid}), representing the non-diagonal terms, goes to zero.
The system tends to become separable \cite{Wer}. It is impossible
to observe such
a behavior in the JC model context.\\
It is to notice that when both the atoms interact with the field
exactly in coincidence of the antinode (i.e. $x_{0}=0$), the
scalar products (\ref{phiupphidownboson}-\ref{phidownphi0boson})
produce just a slow damping of the correlation functions and the
density matrix
remain essentially non-separable.\\
\begin{figure}
 \includegraphics[width = 0.36 \textwidth]{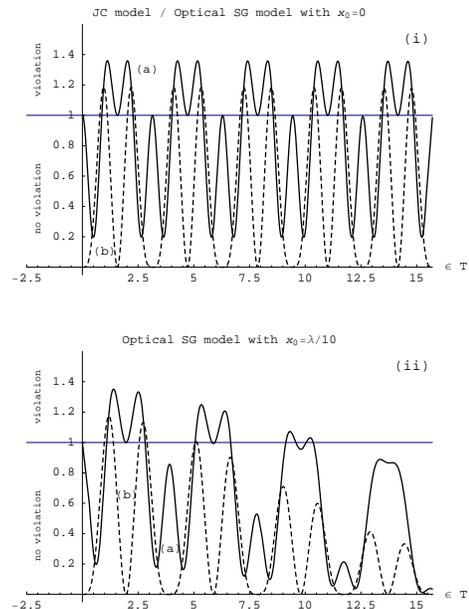}
 \caption{\label{fig1} Graphical solution of Bell's inequality
 in terms of the $M(\rho)=\,max\{(a),(b)\}$ function, for two two-level
 atoms interacting in succession with the field of the same cavity. Figure
 (i) shows the periodicity of $\lambda_{1}+\lambda_{2}$ (continuous
 line) and $2 \lambda_{2}$ (dashed line) when the JC model is
 adopted Ref.\cite{Phoe} or the SG model is considered with
 $x_{0}=0$ for both the atoms.
 Fig. (ii) illustrates the non dissipative damping of the
 correlation between the two atoms
 due to the entanglement of the field and the internal
 atomic variables with the translational atomic degrees of
 freedom when $x_{0} \neq 0$. For little interaction times a magnification of the
 non locality of the entanglement is observed.
 Concerning the translational dynamics, in this graph we suppose
 for both the atoms an initial wave packet of minimum uncertainty, with zero
 mean value of $\hat{p}_{1}$ and $\hat{p}_{2}$, centered in $x_{1}=x_{2}=\lambda/10$ and with
 a width imposed by the condition of dealing with coherent initial
 states. $\lambda = 2 \pi /k$ is the wavelength of the resonant k-mode of the undamped cavity.
 The values of the other parameters are
 $m=10^{-26}$ kg, $\varepsilon=10^8 sec^{-1}$  and  $\lambda=10^{-5}$ meters.}
 \end{figure}

Because, as said above, non-separability does'nt imply a violation
of Bell's inequality, guarantee of non-locality, it is also useful
to investigate the nature of the interatomic correlations in terms
of the Bell's inequality. To this end we consider the Horodecki
family formulation \cite{Hor1}: \emph{A density matrix $\rho$
describing a system composed by two spin $1/2$  subsystems
violates some Bell's inequality in the CHSH formulation} \cite{Cla}
\emph{if and only if the relation $M(\rho)>1$ is satisfied}. The
quantity $M(\rho)$ is defined as follows. Consider the $3\times3$
matrix $T_{\rho}$ with coefficients
$t_{n,m}=tr(\rho\,\sigma_{n}\otimes \sigma_{m})$, where
$\sigma_{n}$ are the standard Pauli matrices. Diagonalizing the
symmetric matrix $U_{\rho}=T_{\rho}^{T}\cdot T_{\rho}$
($T_{\rho}^{T}$ is the transpose of $T_{\rho}$), and denoting the
three eigenvalues of $U_{\rho}$ by $\lambda_{1}$, $\lambda_{2}$
and $\lambda_{3}$, then $M(\rho)=
max\{\lambda_{1}+\lambda_{2},\lambda_{1}+\lambda_{3},\lambda_{2}+\lambda_{3}\}$.
In our case $\lambda_{2}=\lambda_{3}$ and then $M(\rho)=
max\{\lambda_{1}+\lambda_{2},2 \lambda_{2}\}$. Fig. \ref{fig1}
compares the behaviors of $\lambda_{1}+\lambda_{2}$ (continuous
line) and $2 \lambda_{2}$ (dashed line) as a function of the
interaction time for the two models. For simplicity, in both the
figures (i) and (ii) we have assumed $T_{1}=T_{2}=T$. The response
of the Bell's inequality test outlines the great difference
between the interatomic correlations predicted by the two models
when $x_{0} \neq 0$. When $x_{0}=0$ the JC and SG models conduce
to an almost indistinguishable behavior of the system with respect
to non-locality (see Fig. \ref{fig1} (i)). This is due to the fact
that the eq. (\ref{phiupdown1}) reduces to $1/
\sqrt{\cosh(\omega_0 t)}$ when $x_0 = p_0 = 0$ and this term
results slowly decaing for our values of parameters also in
comparison with the decay
in eq. (\ref{phiupdown1red}).\\

It is possible, furthermore, to extend the discussion to another
simple case in which the single excitation belongs initially to
the atom $A_{1}$. For this initial state, the quantity $M(\rho)$
reduces simply to the dashed line of figures (i) and (ii).\\

In conclusion, the internal variables of two atoms that
successively cross an optical cavity may result strongly entangled
through the interaction with the field of the same cavity. This
entanglement may lead to Bell's inequality violation. As it is
known, transfer of information from the system of interest to
other degrees of freedom (to a bath, in the extreme case) produces
a degradation of quantum correlations. For the system here
considered (the optical SG model) the correlation with the atomic
translational degrees of freedom can actually be avoided by
letting the atoms cross as accurately as possible the cavity in
the region with a zero gradient of the mode function.\\
%

% ----------------------------------------------------------------

\end{document}